\documentclass[english,prd,twocolumn,superscriptaddress,nofootinbib,preprintnumbers,10pt]{revtex4}

\usepackage[latin1]{inputenc}
\usepackage{graphicx}
\usepackage{amssymb}
\usepackage{amsfonts}
\usepackage{amsmath}
\usepackage{dcolumn}
\usepackage{hyperref}
\usepackage{babel}

\newcommand{\rd}{{\rm d}}

\def\be{\begin{equation}}
\def\ee{\end{equation}}

\begin{document}


\title{Vainshtein mechanism in Gauss-Bonnet gravity and Galileon aether}

\author{Radouane Gannouji}
\affiliation{Department of Physics, Faculty of Science, Tokyo University of
         Science, 1-3, Kagurazaka, Shinjuku-ku, Tokyo 162-8601, Japan}
\affiliation{Institute of Theoretical Astrophysics, University of Oslo,
             0315 Oslo, Norway}

\author{M. Sami}
\affiliation{Centre of Theoretical Physics, Jamia Millia Islamia, New Delhi-110025, India}

\date{\today}

\begin{abstract}
We derive field equations of Gauss-Bonnet gravity in $4$ dimensions
after dimensional reduction of the action and demonstrate that in
this scenario Vainshtein mechanism operates in the flat spherically
symmetric background. We show that inside this Vainshtein sphere the
fifth force is negligibly small compared to the gravitational force.
We also investigate stability of the spherically symmetric solution,
clarify the vocabulary used in the literature about the
hyperbolicity of the equation and the ghost-Laplacian stability
conditions. We find superluminal behavior of the perturbation of the
field in the radial direction. However, because of the presence of
the non linear terms, the structure of the space-time is modified
and as a result the field does not propagate in the Minkowski metric
but rather in an "aether" composed by the scalar field $\pi(r)$. We
thereby demonstrate that the superluminal behavior does not create
time paradoxes thank to the absence of Causal Closed Curves. We also
derive the stability conditions for Friedmann Universe in context
with scalar and tensor perturbations and we studied the cosmology
of the model.
\end{abstract}

\maketitle


\section{Introduction}
One of the most mysterious discoveries of modern cosmology is
related to late time acceleration of universe
\cite{Riess:1998cb,Perlmutter:1998np}. For past one decade or so,
extensive efforts have been made to understand the underlying
reason of this phenomenon (for a recent review see
\cite{Li:2011sd}). According to the standard lore
, the
acceleration is caused by an exotic form of matter with large
negative pressure called dark energy. The cosmological constant
and a variety of scalar field systems as representative of dark
energy are consistent with observations. Though some of the scalar
field systems with generic features, such as tracker solutions,
are attractive but nevertheless they do not address the conceptual
problems associated with cosmological constant.

There exist an alternative thinking which advocates the need for a
paradigm shift, namely, that gravity is modified at large scales
which might give rise to late time acceleration \cite{Clifton:2011jh}. We know that
gravity is modified at small scales and thus it is quite plausible
that modification also cures at large distance where it has not been
varied directly. As for the small scale corrections, no deviations
from Einstein's theory have been yet detected; perhaps we need to
probe still higher energies to observe these effects. However,
situation is quite different challenging at large scales. Indeed,
Einstein's theory is consistent with observations to a high accuracy
at solar scales thereby telling us that any modification to gravity
should be confronted with tough constraints posed by solar physics.
Secondly, as for the late time acceleration, the modification should
also be distinguished from cosmological constant.

The aforementioned requirements are difficult to satisfy
consistently in f(R) theories of gravity. These theories are
equivalent to GR plus a scalar degree of freedom whose potential is
uniquely constructed from space time  curvature $R$ \cite{DeFelice:2010aj}. The scalar
degree of freedom should mimic quintessence and should hide its
detection at small scales {\it \`a la chameleon} \cite{Khoury:2003aq}.

 The investigations show that the generic $f(R)$ models either reduce to GR plus cosmological
 constant \cite{Thongkool:2009js} or has a $\phi$MDE instead of a standard matter epoch \cite{Amendola:2006we} or hit curvature singularity \cite{Frolov:2008uf} or give
 rise an ugly fine tuning \cite{Faraoni:2011pm} which is the price one has to pay for implementing the
 chameleon mechanism. One might try to remedy these theories
 by complementing them by quadratic curvature corrections \cite{Appleby:2009uf}. However,
 as demonstrated by Lam Hui et al \cite{Hui:2009kc}, theories based upon chameleon
 mechanism lead to a violation of equivalence principal of the order
 of one.

 In view of the aforesaid, we need to look for an alternative
 mechanism of mass screening. The Vainshtein mechanism \cite{Vainshtein:1972sx} is the one
 which gives rise to mass screening: Any modification of gravity
 in the neighborhood of a massive body within a radius dubbed
 Vainshtein radius are switched off kinematically. The mechanism
 was invented to address the discontinuity problem \cite{vanDam:1970vg,Zakharov:1970cc} of massive
 gravity {\it \`a la } Pauli-Fierz \cite{Fierz:1939ix}. In this theory, the zero helicity
 graviton mode $\phi$  is coupled to the stress of energy momentum tensor
 gives rise to serious violations of local gravity constraints
 in the limit of vanishing mass of graviton. Vainshtein pointed out
 that non-linear effects become crucial in this case. The non-linear derivative term added to Pauli-Fierz
 Lagrangian was shown to implement the mass screening thereby
 removing the problem of discontinuity.

 It is interesting to note that the non-linear term
 naturally arises in DGP model \cite{Dvali:2000hr} in the decoupling limit \cite{Luty:2003vm} such that Vainshtein mechanism is in built in the theory. The scalar degree of freedom obeys Galilean
 symmetry in flat space time and is free from ghosts is called {\it
 Galileon} \cite{Nicolis:2008in}. There exist higher order Galileon Lagrangians
in flat and curved space-time \cite{Deffayet:2009wt}. The higher order Lagrangian are
necessary for realizing the late time de-Siter solution \cite{Gannouji:2010au}. The Galileon model appears as a particluar case of the Horndeski action defined in 1974 \cite{Horndeski}. The action has gained some new attention this year, see \cite{Deffayet:2009mn} for more details.

We should also mention that Galileons are deeply related to massive
 gravity \cite{deRham:2010ik} and Dirac-Born-Infeld systems \cite{deRham:2010eu,RenauxPetel:2011dv} in the framework of higher
dimensional theories \cite{Hinterbichler:2010xn,VanAcoleyen:2011mj,Goon:2011qf,Burrage:2011bt}. The spherically symmetric solution is studied in \cite{Kaloper:2011qc}.

In this paper, we consider dimensional reduction of Gauss-Bonnet
gravity \cite{Amendola:2005cr} and look for the possibility of mass screening in the
model. We also investigate causal structure of flat spherically
symmetric and homogeneous and isotropic backgrounds.

\section{Gauss-Bonnet gravity and its Kaluza-Klein reduction}

We shall consider the following action in $D+N$ dimensions,

\be
\mathcal{S}=\int {\rd ^{D+N}\rm{x}}\sqrt{-g^{(D+N)}} \left(R+\alpha R_{\rm GB}\right)
\ee

which is the simplest non trivial form of Lovelock theory.\\

In order to simplify the analysis, we use the metric anzatz,

\be
\rm{ds}^2=g_{\mu\nu}\rm{dx}^\mu\rm{dx}^\nu+e^{\pi}\gamma_{ij}\rm{dx}^i\rm{dx}^j
\ee

where the Greek letters run form $0$ to $D-1$ and the Latin
characters from $D$ to $D+N-1$. The scalar field $\pi$ appearing in
the  metric plays the role of the size of the extra dimensions and depends on the first $D$ coordinates. It
acquires  a non trivial character in case the volume of the
compactified dimensions becomes a variable.

Following a standard prescription for dimensional reduction on a
compact flat space,  we get

\begin{align}
\label{actionD}
& \mathcal{S}=\int {\rd ^{D}\rm{x}}\sqrt{-g^{(D)}} e^{N\pi/2} \Bigl[
R^{(D)}+d_1(\partial \pi)^2+\alpha\Bigl(R_{\rm GB}^{(D)}\nonumber\\
& \hspace{1cm} +d_2G_{\mu\nu}\pi^{;\mu}\pi^{;\nu}+
d_3 (\partial \pi)^4+d_4 (\partial \pi)^2\Box \pi\Bigr)\Bigr]
\end{align}

with
\begin{align}
d_1&=\frac{N(N-1)}{4}\\
d_2&=-N(N-1)\\
d_3&=-\frac{N(N-1)^2(N-2)}{16}\\
d_4&=-\frac{N(N-1)(N-2)}{4}
\end{align}

It should be noticed that the reduced action does not depend on $D$-coefficients.\\

It is interesting to observe that the action in $D+1$ dimensions
reduces to the same form as the original one with a global
factor.

\be \mathcal{S}=\int {\rd ^{D}\rm x}\sqrt{-g^{(D)}}~
e^{\pi/2}\left[R+\alpha R_{\rm GB}\right] \label{action3} \ee

Next, we perform a conformal transformation (in $D$-dimensions) for
transforming the action (\ref{actionD}) to a convenient form. After
rescaling the fields

\be
g_{\mu\nu}\rightarrow \tilde{g}_{\mu\nu}=e^{N\pi/(D-2)}g_{\mu\nu},~~\pi \rightarrow \pm \sqrt{\frac{2(D-2)}{N(D+N-2)}}~\pi
\ee

and carrying out integrations by parts, we obtain,

\begin{align}
&  \mathcal{S}=\int {\rd ^{D}\rm x}\sqrt{-g^{(D)}}\Bigl[R^{(D)}-\frac{1}{2}(\partial \pi)^2+\alpha ~ e^{\beta_D\pi} \Bigl(R_{\rm GB}\nonumber\\
&  \hspace{1cm} +b_1 G_{\mu\nu}\pi^{;\mu}\pi^{;\nu}
+b_2(\partial \pi)^2\Box\pi
+b_3\left(\partial \pi\right)^4\Bigr)\Bigr]\nonumber\\
&  \hspace{1cm} +\mathcal{S}_m[e^{-\beta_D\pi}g_{\mu\nu};\psi_m]
\label{action4}
 \end{align}

where we defined

\begin{align}
b_1&=2\frac{D-2}{D+N-2}\left(1+N\frac{D-4}{(D-2)^2}\right)\\
b_2&=\pm\sqrt{\frac{2}{N}}\left(\frac{D-2}{D+N-2}\right)^{3/2}\left(\frac{N^2}{(D-2)^2}-1\right)\\
b_3&=-\frac{N^2(D-4)+DN(D-2)-2(D-2)^2}{4N(D-2)(D+N-2)}\\
\beta_D &=\pm \sqrt{\frac{2N}{(D-2)(D+N-2)}}
\end{align}

and $\mathcal{S}_m$ is the matter action with matter fields $\psi_m$.\\

We drop in the action the tilde and terms that are total
derivatives. We  notice that the conformal transformation gives the
action an explicit dependance on the dimension $D$.

In $4$-dimensions, the action reduces to

\begin{align}
\label{eq:action}
&  \mathcal{S}=\frac{1}{2}\int {\rd ^{4}\rm x}\sqrt{-g} \Bigl[R-\frac{1}{2}(\partial \pi)^2
+\alpha  e^{\beta\pi} \Bigl(R_{\rm GB}\nonumber\\
&  \hspace{1cm} +c_1G_{\mu\nu}\pi^{;\mu}\pi^{;\nu}+c_2(\partial \pi)^2\Box\pi+c_3\left(\partial \pi\right)^4\Bigr)\Bigr]\nonumber\\
&  \hspace{1cm} +\mathcal{S}_m[e^{-\beta \pi}g_{\mu\nu};\psi_m]
\end{align}

with
\begin{align}
\beta\equiv \beta_4 &=\pm \sqrt{\frac{N}{N+2}},\\
c_1&=\frac{4}{N+2},\\
c_2&=\frac{N-2}{N}\beta,\\
c_3&=-\frac{N-1}{N(N+2)}.
 \end{align}
In what follows, we shall investigate the dynamics based upon the
action (\ref{eq:action}).

\section{Field equations of motion}

The field equations one derives from the action (\ref{eq:action}) are

\begin{align}
\label{EM}
& \Box \pi+\alpha e^{\beta\pi} K=\beta T,\\
\label{EM2}
& G_{\mu\nu}-\frac{1}{2}\left(\pi_{;\mu}\pi_{;\nu}-\frac{1}{2}g_{\mu\nu}(\partial \pi)^2\right)+\alpha e^{\beta \pi}\Sigma_{\mu\nu}= T_{\mu\nu}.
\end{align}

with

\begin{align}
K&=
\beta R_{\rm GB}
-c_1 G_{\mu\nu}\left(\beta\pi^{;\mu}\pi^{;\nu}+2\pi^{;\mu\nu}\right)
-4c_3(\partial \pi)^2\Box\pi\nonumber\\
&+4\left(\beta c_2-2c_3\right)\pi^{;\mu\nu}\pi_{;\mu}\pi_{;\nu}
+\beta\left(\beta c_2-3c_3\right)(\partial\pi)^4
\nonumber\\
&-2c_2\left[(\Box\pi)^2
-\pi^{;\mu\nu}\pi_{;\mu\nu}
-R_{\mu\nu}\pi^{;\mu}\pi^{;\nu}\right],\\
{\Sigma}_{\mu\nu}&=
-2\left(\beta\pi^{;\rho\sigma}+(\beta^2+\frac{c_1}{4} )
\pi^{;\rho}\pi^{;\sigma}\right)\Bigl[2R_{\rho\mu\nu\sigma}+2\Bigl(g_{\mu\nu}R_{\rho\sigma}\nonumber\\
&+R_{\mu\nu}g_{\rho\sigma}-2g_{\rho(\mu}R_{\nu)\sigma}\Bigr)+R(g_{\rho\mu}g_{\sigma\nu}-g_{\rho\sigma}g_{\mu\nu})\Bigr]
\nonumber\\
&+\frac{1}{2}g_{\mu\nu}\left(\beta c_2-c_3\right)(\partial\pi)^4
+\left(2c_3-\beta c_2\right)(\partial\pi)^2\pi_{;\mu}\pi_{;\nu}
\nonumber\\
&+\left(c_2-\frac{\beta}{2}c_1\right)\Bigl[\pi_{;\mu}\pi_{;\nu}\Box \pi
-2\pi^{;\sigma}\pi_{;\sigma(\mu}\pi_{;\nu)}\Bigr.\nonumber\\
&
\Bigl.
+g_{\mu\nu}\pi^{;\rho}\pi_{;\rho\sigma}\pi^{;\sigma}\Bigr]
+c_1\Bigl[
\pi_{;\mu\rho}\pi^{;\rho}_{~\nu}
-\pi_{;\mu\nu}\Box\pi
+\frac{1}{2}G_{\mu\nu}(\partial\pi)^2\nonumber\\
&+\frac{1}{2}g_{\mu\nu}\left((\Box\pi)^2-\pi_{;\rho\sigma}\pi^{;\rho\sigma}\right)
+\frac{\beta}{2}(\partial\pi)^2\Bigl(g_{\mu\nu}\Box\pi-\pi_{;\mu\nu}\Bigr) \Bigr].
\end{align}

In the analysis to follow, we shall focus on equations of motion
(\ref{EM}) to study mass screening induced by non-linear terms in a
simple tractable background.
\section{Mass screening $-$ Vainshtein mechanism}

In order to investigate the effects of the non linear terms in the
action, we shall study the model
in a flat spherically symmetric background. \\
In this case, the theory reduces to a special case of KGB
\cite{Deffayet:2010qz} coupled to matter,

\be
\mathcal{S}=\frac{1}{2}\int {\rd ^{4}\rm x} \left[K(\pi,X)+G(\pi,X)\Box\pi\right]+\mathcal{S}_m[e^{-\beta \pi}g_{\mu\nu};\psi_m]
\ee

with

\begin{align}
&K(\pi,X)=X-4\alpha\frac{N-1}{N(N+2)}e^{\beta\pi}X^2\\
&G(\pi,X)=-2\alpha\beta\frac{N-2}{N}e^{\beta\pi}X,~\text{and}~X=-\frac{(\partial\pi)^2}{2}
\end{align}

And the special case of $N=2$ gives rise to K-essence.

The equation of motion is

\begin{align}
&\pi''+2\frac{\pi'}{r}+\alpha e^{\beta\pi}\frac{\pi'}{N(2+N)r^2}\Bigl[
8(N-1)r\pi'^2\nonumber\\
&\hspace{1cm}+(N^2+N-3)r^2\pi'(\beta \pi'^2+4\pi'')\nonumber\\
&\hspace{1cm}-4\beta(N^2-4)(\pi'+2r\pi'')\Bigr]=-\beta \rho
\end{align}

where $'$ represents the derivative with respect to r.\\
If we integrate this equation from $r=0$ to a distance outside the
body in the case of $\alpha=0$, we have $\pi'=-\frac{\beta
r_s}{r^2}$ (where $r_s$ is the Schwarzschild radius $r=2G M$) and the fifth force is of the order of the gravitational
force, \begin{align}
\left|\frac{\mathcal{F}_\pi}{\mathcal{F}_g}\right|=\frac{\beta
r^2}{r_s}|\pi'|=\frac{N}{N+2} \simeq 1\end{align}

Let us note that we have two different values of $\beta$; it
is easy to see that if we change $\beta \rightarrow -\beta$ the action remains unchanged provided that,
 $\pi \rightarrow -\pi$. Therefore the fifth force is invariant
 under the change of sign of $\beta$.\\

Unfortunately we can not get analytical solutions in case  $\alpha
\neq 0$.  However,  we can have derive asymptotic solutions at large
and short distances.

At large distances, we observed that the solution is trivial. This
solution should change as we approach the source of matter because
of the effect of the non linear terms. Therefore these corrections
to the asymptotic solutions become crucial when we enter  the
Vainshtein radius.

We define this scale as the radius where a perturbation of this
trivial solution becomes important. It is easy to show that the
Vaishtein radius can be approximated by

\begin{align}
\text{For $N=2$,  } R_V^4\simeq \alpha r_s^2\\
\text{For $\forall$ $N\neq 2$,  } R_V^3\simeq \alpha r_s
\end{align}

\begin{figure}
\includegraphics[scale=.7]{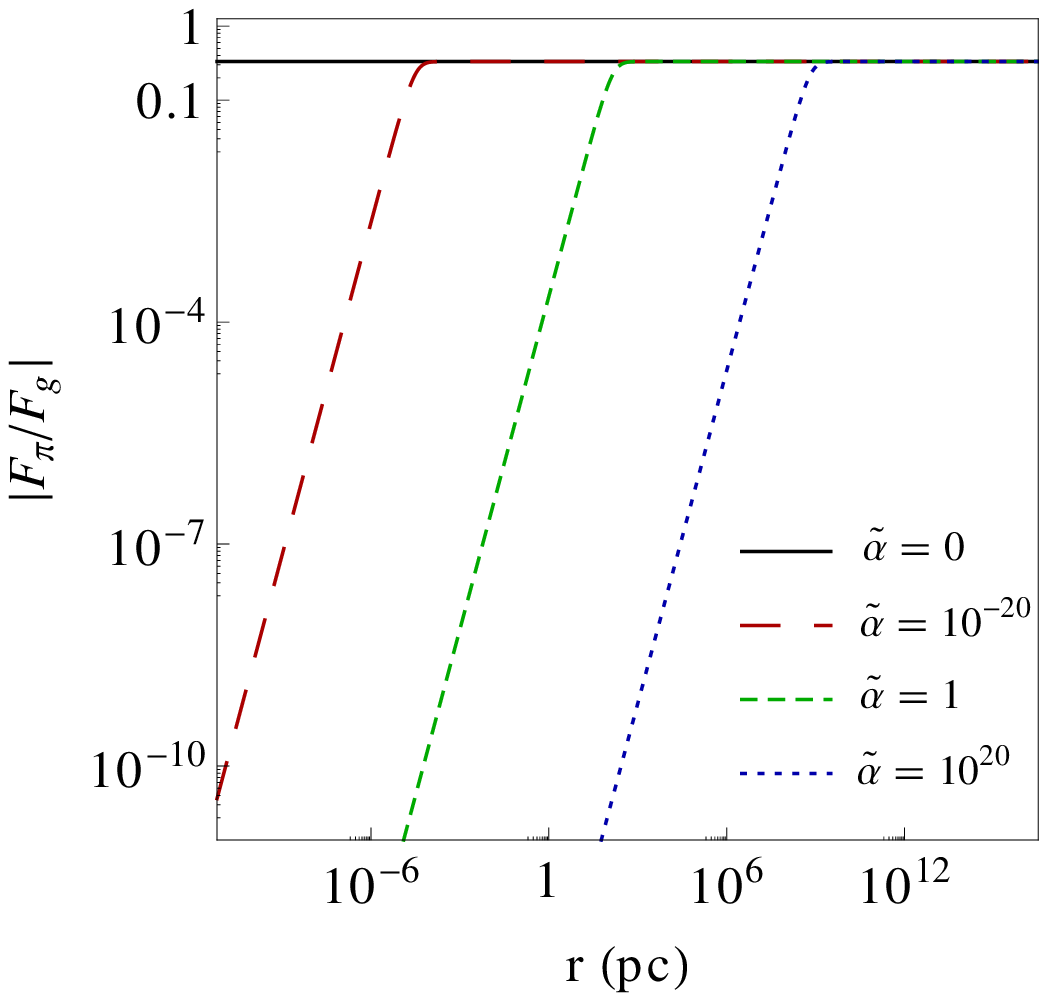}
\includegraphics[scale=.7]{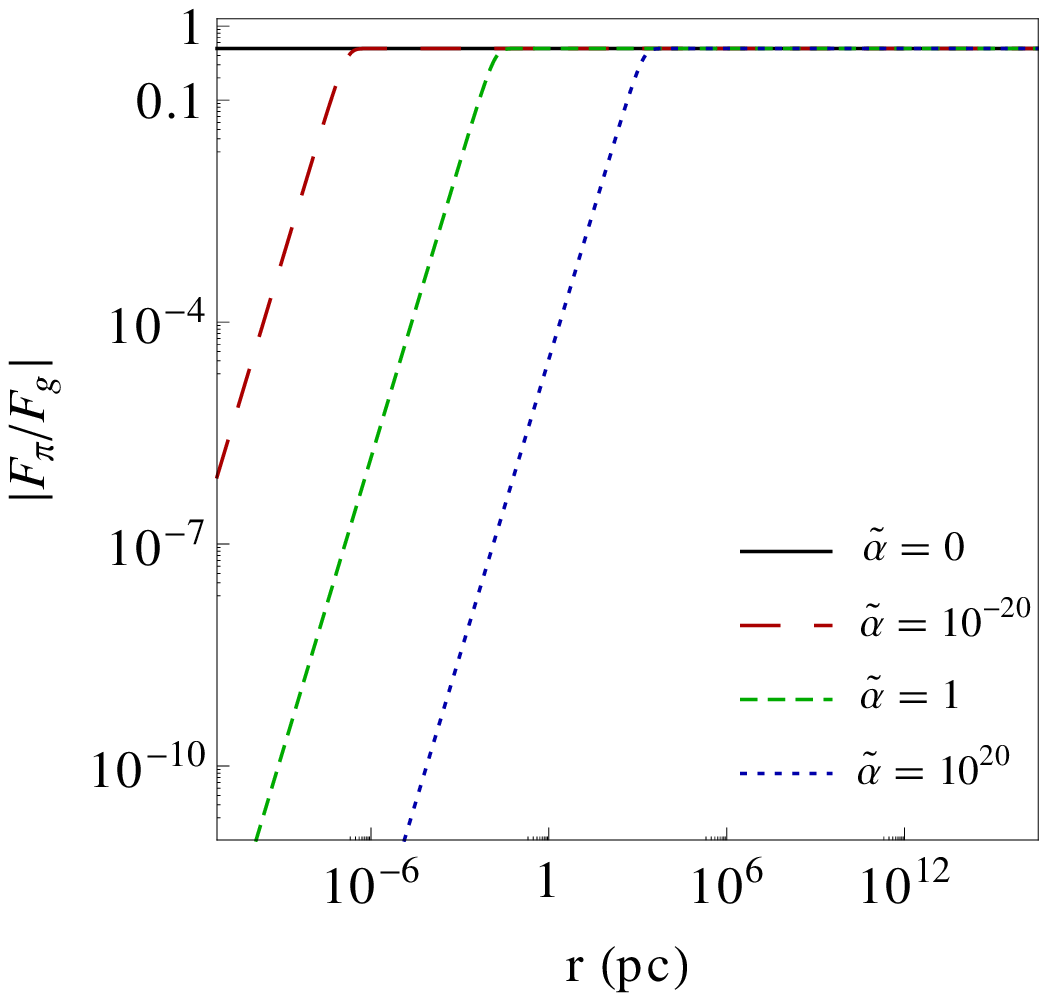}
\caption{(Top):The ratio of the fifth force and the gravitational force versus the distance from the source in parsecs, for $N=1$.
In the numerics we considered $r_s\equiv r_s$(Sun) and $\bar \alpha\equiv H_0^2\alpha$.
(Bottom):The same evolution for $N=2$. The fifth force is negligeable at small distances compared to the gravitational force.}
\label{fig:LGC}
\end{figure}

In case,  $\sqrt \alpha$ is of the order of the Hubble scale
$(\tilde \alpha\equiv H_0^2 \alpha=1)$, we have for $N=1$ $R_V\simeq
10^2 $pc and $R_V\simeq 2.10^{-2}$pc for $N=2$, which is in perfect
agreement with the Fig.\ref{fig:LGC}, where we considered the Sun as
the central body.

In the Fig.\ref{fig:LGC}, we show that for $N=1$ and $N=2$ there is
  a screening effect at distances smaller that the Vainshtein radius.
 For dimensions larger than $2$, the evolution of the fifth force
 versus the radial coordinate is the same as in case of $N=1$. In fact
 for $N=2$ we don't have
the G-Essence term\footnote{The extra term in the Lagrangian
compared to K-Essence: $G(\pi,X)$} which gives an additional effect
to the screening mechanism as it can be seen in the
Fig.\ref{fig:LGC}.

We also find from  numerical analysis, that for the solution to
be continuous, we need for $N\leq 2$ in case of $\alpha>0$ and $\alpha<0$ for $N>2$.\\

It was shown in \cite{Amendola:2007ni} that we have extremely tight bounds on
the parameters of the model, but as we saw, if we consider the full-action without
any approximation on the non-linear terms, we have a Vainshtein mechanism which
allows the coupling $\alpha$ to take large values.

\section{Stability of solutions}

We consider the test field approximation where we expand the field
$\pi\rightarrow \pi+\phi$ and neglecting the back reaction on the
metric. The equation for the scalar field ($\phi$) in the first
order are given by,

\be
\label{effective}
G^{\mu\nu}_{\text{eff}}\phi_{;\mu\nu}+V^{\mu}\phi_{;\mu}+M\phi=0
\ee

where the induced metric is

\be
G^{\mu\nu}_{\text{eff}}=A g^{\mu\nu}+B\pi^{;\mu} \pi^{;\nu} +C \pi^{;\mu\nu}
\ee

with

\begin{align}
A&=1+\frac{4\alpha}{N} e^{\beta \pi}\left(\frac{N-1}{N+2}\pi'^2-\beta(N-2)\left(\pi''+\frac{2}{r}\pi'\right) \right)\nonumber\\
B&=4 \frac{N^2-2}{N(N+2)}\alpha e^{\beta \pi}\nonumber \\
C&=4\beta\frac{N-2}{N}\alpha e^{\beta \pi}
\end{align}

The field equation admits a well-posed initial value-formulation locally if the
effective metric $G^{\mu\nu}_{\text{eff}}$ is Lorentzian.

The equation (\ref{effective}) can be expanded as

\be
\label{effective2}
G^{00}_{\text{eff}}\partial_t^2\phi
+G^{11}_{\text{eff}}\partial_r^2\phi
+G^{22}_{\text{eff}}r^2\partial_\Omega^2 \phi
+\text{~first derivatives of }\phi+\cdots =0
\ee

where $\partial_\Omega^2 \phi$ is the angular part of the Laplacian.

This equation is Lorentzian with a signature (-,+,+,+), if we
have $G^{00}_{\text{eff}}<0$, $G^{11}_{\text{eff}}>0$ and $G^{22}_{\text{eff}}>0$.

These conditions are exactly the same as  the ghost free condition
and the stability
of the Laplacian used in the literature.\\

In fact the equation (\ref{effective2}) can be derived from the action

\be
\delta^2 S=\int -\frac{1}{2}G^{00}_{\text{eff}}\left[\left(\partial_t\phi\right)^2-c_r^2\left(\partial_r\phi\right)^2-c_\Omega^2\left(\partial_\Omega\phi\right)^2\right]{\rm d}^4 x
\ee

where

\begin{align}
c_r^2&=-\frac{G^{11}_{\text{eff}}}{G^{00}_{\text{eff}}} =1+\frac{B}{A}\pi'^2+\frac{C}{A}\pi''\\
c_\Omega^2 &=-r^2\frac{G^{22}_{\text{eff}}}{G^{00}_{\text{eff}}}=1+\frac{C}{rA}\pi'
\end{align}

The ghost condition fixes the
 sign of $G^{00}_{\text{eff}}<0$ and $G^{11}_{\text{eff}}>0$ and $G^{22}_{\text{eff}}>0$ via
 the positivity of the sound speed $c_r^2>0$, $c_\Omega^2>0$( also know as the stability of the Laplacian).\\

When the non linear terms ($\alpha$-terms) are dominant, we can
reduce the evolution equation for the scalar field to

\begin{align}
&N=2,~~2\pi'+3r\pi''=0 \Rightarrow \pi'\propto r^{-2/3},\\
&N\neq 2,~~\pi'+2 r \pi''=0 \Rightarrow \pi' \propto r^{-1/2}.
\end{align}

Thus it becomes easy to estimate the sound speed at small distances,

\begin{align}
&N=2,~~c_r^2=3,~~c_\Omega^2 =1\\
&N\neq 2,~~c_r^2=4/3,~~c_\Omega^2 =1/3.
\end{align}

The propagation of the perturbation of the scalar field is therefore
superluminal in the radial direction $c_r^2>1$ at small distances.
However, we shall demonstrate  in the next section for the special
case $N=2$ that the superluminal does not implies a non causal
propagation of the perturbations. It is in fact just a redefinition
of the maximum sound speed via a larger light cone structure of the
space time compared to the Minkowsky space.

\section{Special case of $N=2$}

In the particular case of $N=2$ where an emergent geometry is
present (see \cite{Babichev:2007dw} for more details), we have

\be
c_r^2=\frac{1-3\alpha X e^{\beta\pi}}{1-\alpha X e^{\beta\pi}}
\ee

Hence as soon as the non linear  $(\alpha)$ term becomes dominant,
we have $c_r^2 \propto 3$.

\begin{figure}
\includegraphics[scale=.7]{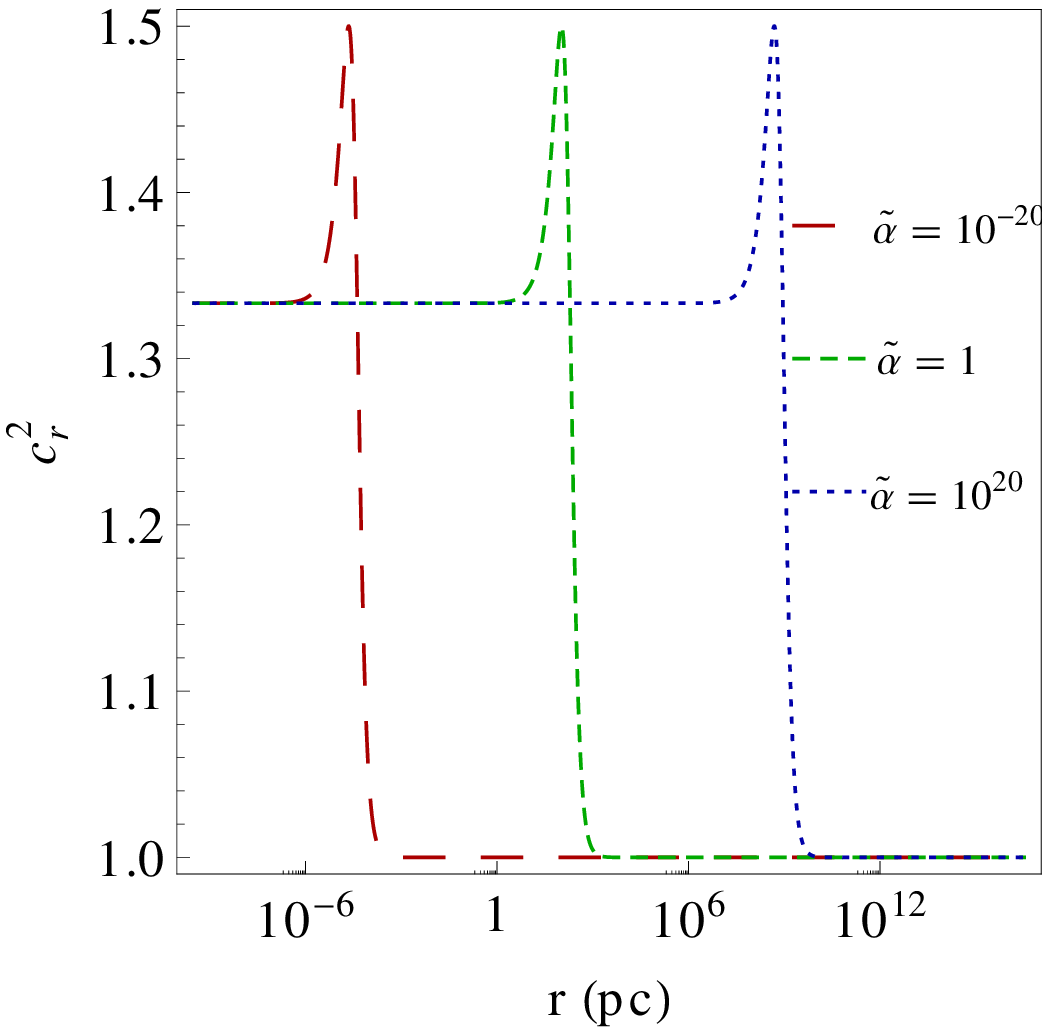}
\includegraphics[scale=.7]{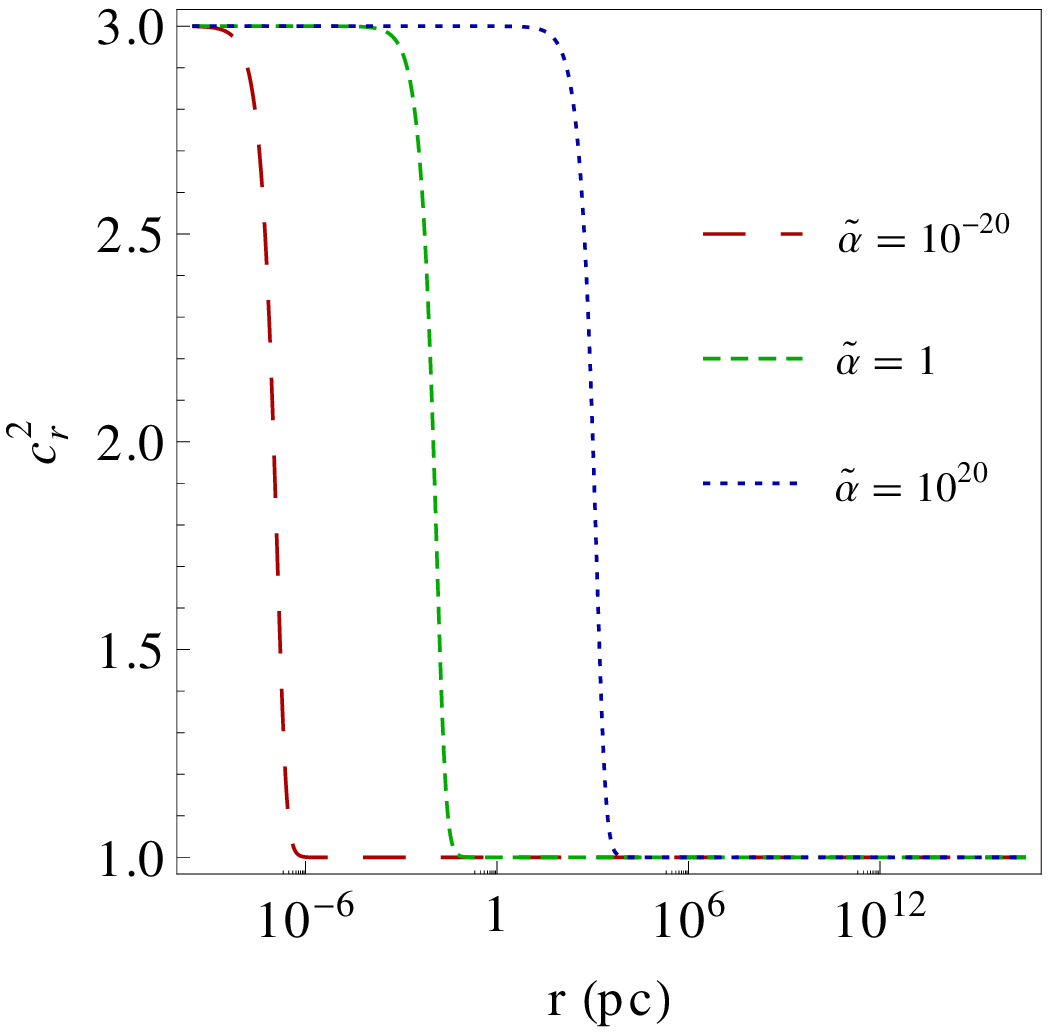}
\caption{(Top): The radial sound speed versus the distance in parsecs for $N=1$, (Bottom): The same figure for $N=2$}
\end{figure}

\begin{figure}
\includegraphics[scale=.7]{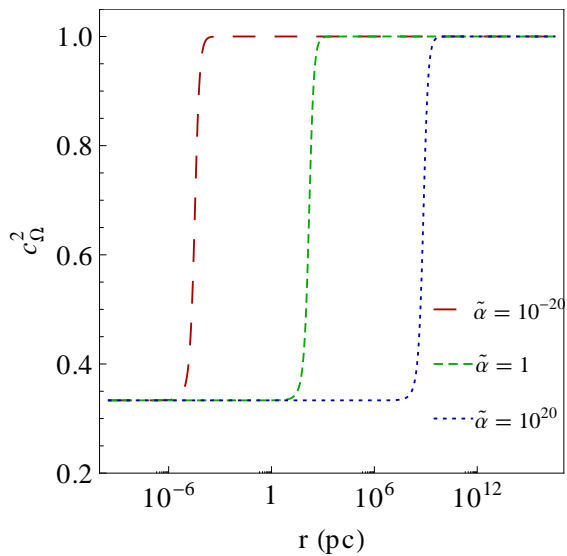}
\caption{The angular sound speed versus the radial distance in parsecs for $N=1$}
\end{figure}

The non linear terms which are necessary for local constraints
create automatically a superluminal propagation. This behavior is
present in models involving Galileons and their extensions \cite{Nicolis:2008in,Andrews:2010km,Goon:2010xh}.

 We should, however, emphasize that we do not have this
propagation in the Mikowski space-time
but in an extended structure of space-time because of the non linear terms.\\
In fact, in the special case of $N=2$, the model reduces to a
particular K-essence; therefore by performing a conformal
transformation we have,

\be
\tilde{G}^{\mu\nu}_{\text{eff}}=\frac{1}{K_{,X}^2c_r} G^{\mu\nu}_{\text{eff}}
\ee

We can rewrite the equation for perturbations as \cite{Babichev:2007dw}

\be
\tilde{G}^{\mu\nu}_{\text{eff}}D_{\mu}D_{\nu}\phi-M^2_{\text{eff}}\phi=0
\ee

where $D_\mu$ is the covariant derivative associated to the effective metric $\tilde{G}^{\mu\nu}_{\text{eff}}$.

and

\be
M_{\text{eff}}^2=-\frac{\alpha}{K_{,X}^2c_r}\left(\frac{3}{4}X^2-\beta X \Box \pi+\beta \pi^\mu\pi_{\mu\nu}\pi^\nu\right)e^{\beta\pi}
\ee

We  notice the difference with \cite{Babichev:2007dw} where the
scalar field was time dependant \footnote{In fact, if $\pi\equiv
\pi(t)$, the sound speed in an homogenous and isotropic Universe is
$c_s^2=(1-2X \frac{K_{,XX}}{K_{,x}} )^{-1}$ but in the case where
$\pi\equiv \pi(r)$ the sound speed in the radial direction is
$c_r^2=1+2X \frac{K_{,XX}}{K_{,x}}$.}.

It is also interesting to note that the mass term is null when
$\alpha \rightarrow \infty$.

Therefore the emergent geometry defined by the metric
$\tilde{G}^{\mu\nu}_{\text{eff}}$ defines a new structure of the
space-time, the light cone is larger than the standard one if and
only if $\frac{K_{,XX}}{K_{,X}}<0$. In the limit where the non
linear terms are dominant, we have
$\frac{K_{,XX}}{K_{,X}}=-\frac{2}{\pi'^2}<0$.

We emphasize also that this new structure of the spacetime is stably
causal. In fact the Minkowski time could define a future directed
timelike vector field,

\be
\tilde{G}^{\mu\nu}_{\text{eff}}\partial_{\mu}t\partial_{\nu}t=-\frac{1}{K_{,X}c_{r}}
\ee

which is negative as soon as the hyperbolicity conditions are satisfied.\\
Therefore no CCCs (Causal Closed Curves) can exist.

The superluminal propagation does not create any causal
inconsistencies. In fact the perturbations of the scalar field do
not propagate in the Minkowski space-time but rather in some form of
"aether" because of the presence of the background field $\pi(r)$.
The maximum of the speed of the field is just a redefinition of the
speed of light in this new space-time. The causal structure is not
changed, in the sense that we do not have CCCs in this case.

\section{Background cosmological dynamics}

We concentrate on spatially flat
Friedman-Lema\^itre-Robertson-Walker (FLRW) universes with  scale
factor $a(t)$,

\be
{\rm d s}^2=-{\rm d t}^2+a^2 {\rm d \textbf x}^2
\ee

\begin{align}
\label{eq:F1}
\ddot\pi+3H\dot\pi-\alpha e^{\beta\pi}K&=\beta\rho,\\
\label{eq:F2}
3H^2-\frac{1}{4}\dot\pi^2-\alpha e^{\beta\pi}\Sigma^{0}_{~0}&=\rho,\\
\label{eq:F3}
3H^2+2\dot H+\frac{1}{4}\dot\pi^2-\alpha e^{\beta\pi}\Sigma^{1}_{~1}&=0.
\end{align}

The following equations are obtained

\begin{align}
&\Sigma^{0}_{~0}=
-12\beta H^3\dot\pi
+\frac{9}{2}c_1H^2\dot\pi^2
-\frac{c_2}{2}\dot\pi^3\left(\beta\dot\pi-6H\right)\nonumber\\
&\hspace{1cm}+\frac{3}{2}c_3\dot\pi^4,\\
&\Sigma^{1}_{~1}=
-4\left[(\beta \ddot\pi+\beta^2\dot\pi^2+2\beta H\dot\pi)H^2+2\beta H\dot H\dot\pi \right]\nonumber\\
&\hspace{1cm}+\frac{1}{2}c_1\dot\pi\left[\dot\pi(2\dot H+3H^2+2\beta H\dot\pi)+4H\ddot\pi\right]\nonumber\\
&\hspace{1cm}+\frac{1}{2}c_2\dot\pi^2(2\ddot\pi+\beta\dot\pi^2)
-\frac{1}{2}c_3\dot\pi^4,\\
&K=24\beta H^2(H^2+\dot H)
-3c_1\bigl[H^2(\beta\dot\pi^2+2\ddot\pi+6H\dot\pi)\nonumber\\
&\hspace{1cm}+4H\dot H\dot\pi \Bigr]
+c_2\dot\pi\Bigl[\dot\pi(4\beta\ddot\pi+\beta^2\dot\pi^2-6\dot H-18H^2)\nonumber\\
&\hspace{1cm}-12H\ddot\pi\Bigr]
-3c_3\dot\pi^2\left[\beta\dot\pi^2+4H\dot\pi+4\ddot\pi\right].
\end{align}

where $H\equiv \frac{\dot a}{a}$ is the Hubble rate while a dot
stands for a derivative with respect to the cosmic time $t$.

\section{Stability conditions in an FLRW Universe}

In order to derive the stability conditions of the theory in the
context of an isotropic and homogeneous Universe, we study  linear
perturbation in a FLRW background. We consider the following metric

\begin{align}
&{\rm d s}^2=-\left(1+2\alpha\right){\rm d t}^2-a\beta_{,i}{\rm d t} {\rm d x}^i\nonumber\\
&\hspace{.8cm}+a^2\left[\delta_{ij}(1+2\psi)+2\gamma_{;ij}+2h_{ij}\right]{\rm d x}^i{\rm d x}^j.
\end{align}

where $\alpha$, $\beta$, $\psi$ and $\gamma$ are scalar metric perturbations,
and $h_{ij}$ is the traceless and divergence-free tensor perturbations.

We did not considered vector perturbations in the line element because of the absence of an anisotropic fluid.\\

\subsection{Scalar perturbations}

For scalar perturbations, we can neglect the matter contributions at late times, and it was noticed in
 that the calculations simplify if we  work in the uniform-field gauge $\delta\pi=0$ \cite{Hwang:1999gf}.\\

Therefore we can show that the action at the second order can be written in the following form

\be
\delta^2 S=\frac{1}{2}\int {\rm d x}^3 {\rm d t}a^3 Q^{(s)}\left[\dot\psi^2-\frac{c_s^{2(s)}}{a^2}(\partial_i \psi)^2\right]
\ee

where
\begin{align}
   & Q^{(s)} = \frac{\dot\pi^2
       +6\frac{\left(Q^{(s)}_{a}\right)^2}{2
       +Q^{(s)}_{b}}+2Q^{(s)}_{c}}
       {\left(H+\frac{Q^{(s)}_{a}}{2
       +Q^{(s)}_{b}}\right)^{2}},
   \nonumber \\
   & c_s^{2(s)} = 1 + 2\frac{Q^{(s)}_{d}
       +\frac{Q^{(s)}_{a}Q^{(s)}_{e}}{2
       +Q^{(s)}_{b}}-\left(\frac{
       Q^{(s)}_{a}}{2+Q^{(s)}_{b}}\right)^2 Q^{(s)}_{f}}
       {\frac{\dot\pi^2}{2}+3\frac{\left(
       Q^{(s)}_{a}\right)^2}{2+Q^{(s)}_{b}}+Q^{(s)}_{c}}.
   \label{e:s_scalar}
\end{align}

where we defined

\begin{align}
& Q^{(s)}_{a} = \alpha \left[ 4 \beta H^2 - 2 c_1 \dot \pi H
    - c_2 \dot \pi^2 \right]\dot \pi e^{\beta\pi},\\
& Q^{(s)}_{b} = \alpha\left[ 8 \beta H - c_1 \dot \pi \right]\dot \pi e^{\beta\pi},\\
& Q^{(s)}_{c} = \alpha \left[ 3 c_1 H^2 + 2 c_2 \dot \pi \left( 3 H
    - \beta \dot \pi \right) + 6 c_3 \dot \pi^2 \right] \dot \pi^2 e^{\beta\pi},\\
& Q^{(s)}_{d} = \alpha \Big[c_1 \dot H + c_2 \left( \beta \dot \pi^2
    + \ddot \pi - \dot \pi H \right)- 2 c_3 \dot \pi^2 \Big] \dot \pi^2 e^{\beta\pi},\\
& Q^{(s)}_{e} = \alpha \Big[ 8 \beta \dot H - c_1
    \left( \beta \dot \pi^2 + 2 \ddot \pi- 2 \dot \pi H \right)
    + 2 c_2 \dot \pi^2 \Big]\dot \pi e^{\beta\pi},\\
& Q^{(s)}_{f} = \alpha \left[ 4 \left( \beta \ddot \pi+\beta^2\dot\pi^2 -\beta \dot \pi H
    \right)+ c_1 \dot \pi^2 \right] e^{\beta\pi}.
\end{align}

We recover the results derived in \cite{Cartier:2001is}, see also \cite{Kobayashi:2011nu} for a generalization of the model in the context of inflation.\\

The ghost condition fixes the sign of $Q^{(s)}>0$, and also the
positivity condition of the sound speed fixes $c_s^{(s)}>0$. As we
mentioned before that hyperbolicity of equations of motion implies
that we have a well posed Cauchy problem (at least locally).

\subsection{Tensor perturbations}

We can also show that for the tensor perturbations we have

\begin{align}
& {1 \over a^3 Q^{(t)}} \left( a^3 Q^{(t)} \dot{h}^{i}_{j}
     \right)^\cdot - c_s^{(t)} {\Delta \over a^2} h^{i}_{j}
      = {1 \over Q^{(t)}} \delta T^{(t)i}_{~~~~j},
\end{align}
where
\begin{align}
& Q^{(t)} = 2 + \alpha
    \left( 8 \beta H -c_1 \dot{\pi} \right)\dot\pi e^{\beta\pi},\nonumber \\
& c_s^{(t)} = {1 \over Q^{(t)}}
    \left[ 2 + \alpha \left( 8 \left(\beta\ddot{\pi}+\beta^2\dot\pi^2\right) + c_1 \dot{\pi}^2
    \right)e^{\beta\pi} \right].
\end{align}

Similar to the case of scalar perturbations, we have the two
conditions of stability,

 \be Q^{(t)}>0,~~~ c_s^{(t)}>0 \ee

\subsection{Cosmology of the model}

In this section we study the cosmology of the model reduced
in $4$-dimensions (\ref{eq:action}). By considering the following variables
\begin{align}
x&=8\alpha H^2 e^{\beta \pi}\,,\\
y&=\frac{\beta}{2}\frac{\dot \pi}{H}
\end{align}
The Friedmann equations (\ref{eq:F1},\ref{eq:F2},\ref{eq:F3}) reduce
to a very simple form which depend only on the 3 variables
$(x,y,\Omega_r)$, and the dimension $N$,  where $\Omega_r$
corresponds to the radiation. The autonomous system does not depend
on the coupling constant $\alpha$. The relevant fixed points are,
the radiation phase $(x,y,\Omega_r)=(0,0,1)$; there is also a point
which corresponds to an accelerated Universe, namely,
$(x,y,\Omega_r)=(-2\frac{(2+N)^2}{N(1+N)},\frac{N}{2+N},0)$, for
which we have $\Omega_m=0$ and $w_{\rm eff}=-\frac{6+N}{6+3N}$. we
recover a de Sitter phase in the case, $N=0$.
 For all values of $N$, this point corresponds to an accelerated phase of the Universe. We did not find a matter
 phase in the model under consideration. The closest fixed point corresponds to $(x,y,\Omega_r)=(0,\frac{N}{2+N},0)$ for which we
 have $\Omega_m=\frac{2(2+N)}{3(2+N)}$ and $w_{\rm eff}=\frac{N}{6+3N}$, in the limit $N\rightarrow
 0$,
 we recover a standard matter phase. We emphasize that between the "matter" phase and the
 accelerated phase $y=\frac{N}{2+N}$ is constant, the system has a tracker solution.\\
Unfortunately the model is not viable because of the absence of a
standard matter phase. It seems that the reduction of the model in
$4$-D is not pertinent at large scales relevant to cosmological
dynamics . In this scenario, cosmology could be studied in
$4+D$-dimensions. On the other hand, it is known that various
possible  reduction schemes from higher to 4-dimensions
 can give rise to a potential term which could give rise to a viable
 cosmology. We defer these investigations  to our future project.

\section{Conclusion and Perspectives}

In this paper we have studied simple extension of General Relativity
in the context of Lovelock theory. We derived the equations of the
reduced theory in $4$ dimensions. We have shown that locally in a
flat spherically symmetric background,  the non linear terms coming
from the Gauss-Bonnet term in higher dimensions induce Vainshtein
mechanism. We found that in $4+2$-dimensions, the Vainshtein radius
can be approximated by, $R_V^4\simeq \alpha r_s^2$ whereas in case
of $4+N$-dimensions by $R_V^3\simeq \alpha r_s$ with $N\neq 2$.
 We have shown that at distances lower than the Vainshtein radius, the fifth force is negligibly small
 compared to the gravitational force.\\
We have investigated the behavior of the scalar field inside the
Vainshtein sphere and  derived stability conditions of the model. We
have reaffirmed that the hyperbolicity of the equations is
equivalent to the ghost condition and the stability of the
Laplacian. We  found that the model has a superluminal propagation
of the perturbation of the scalar field in the flat spherically
symmetric solution. This faster than light solution appears
 as soon as the non linear terms of the model become dominant. We have shown, in the special case
 of $N=2$ that the causality structure of the space time is well defined even in the
 presence of the superluminal propagation. In fact, we shown that the field propagates
  in a space-time which is not anymore the Minkowski one but some kind of "aether"
   because of the presence of the background field $\pi(r)$. This modification of
   the structure of the space-time is related to the domination of
   the non linear terms in the Lagrangian. We observed that the light cone gets
   wider
   in the aether as compared to the case of Minkowski spacetime provided that the stability conditions
   hold thereby demonstrating that no CCCs appears even in the
   presence of the superluminal propagation.\\
Finally in the context of an isotropic and homogeneous Universe, we
derived Friedmann equations for the field and established the
stability conditions in the context of the scalar and tensor
perturbations. We show that the matter phase is absent in
the  model under consideration.

It will be interesting to investigate the cosmological dynamics and
observational constraints on the model under consideration,
in the presence of a potential term, in a separate
publication. It is also important to investigate the model with
general Lovelock terms in simple and non-trivial topology of extra
dimensions. We defer this work to our future investigations.

\section*{ACKNOWLEDGEMENTS}
The work of RG was supported by the Grant-in-Aid for Scientific Research Fund of the JSPS No. 10329

\end{document}